\documentclass[aps,prb]{revtex4}
\usepackage{amsmath}
\usepackage{graphicx}

\begin{document}

\title{Electric double layer effect  
in an extreme  near-field heat transfer between metal surfaces}

\author{A.I. Volokitin$^{*}$}

\affiliation{
Samar State Technical University,  443100 Samara, Russia}

\begin{abstract}
Calculations of heat transfer between two plates of gold in an extreme near field is performed taking into account the existence of the electric double layer on metal surfaces. For $d<3$nm the double layer contribution exceeds the predictions of the conversional theory of the heat transfer by several orders of magnitudes. This effect is due to  a  coupling  between the radiation electric field and   the double layer dipole moment. The results obtained  can be used for the heat management at the nanoscale by intentionally changing the parameters of surface dipoles with the help of engineering.  
\end{abstract}
\maketitle


\vskip 5mm

 All bodies are surrounded by a fluctuating electromagnetic field due to quantum and thermal fluctuations. This electromagnetic field is responsible for  the fluctuation-induced electromagnetic phenomena (FIEP), such as  the Casimir force, that  play an extremely important role in nanotechnology because it is one of the dominant forces at the nanoscale.  At present a great deal of attention is attracted for  studies of  FIEP at    dynamic and thermal nonequilibrium conditions. Under these conditions, a modification of the Casimir force appears and new effects arise such as the radiative heat transfer and Casimir friction, and it becomes possible to control  FIEP, which is extremely important for the development of micro- and nanoelectromechanical devices. FIEP are usually described by  a fluctuation electrodynamics developed by Rytov\cite{Rytov1953,Rytov1967,Rytov1987}. In the framework of this theory it was theoretically predicted
\cite{Polder1971PRB,Pendry1999JPCM,Greffet2005SSR,Volokitin2001PRB,Volokitin2007RMP}  and experimentally confirmed  \cite{Shen2009NanoLett,Greffet2009NatPhot,Reddy2015AIP,Kittel2017NatCommun,Reddy2017NatCommun,Reddy2015Nature}, that the radiative heat flux between two bodies with different temperatures in the near field (when the distance between the bodies   $d<\lambda_T=c\hbar/k_BT$: at room temperature $\lambda_T\sim 10\mu$m) can be by many orders of magnitude larger than the limit, which is established by Planck's law for blackbody radiation. With the development of new experimental techniques over the past decade, super-Planckian heat transfer has been observed for vacuum gaps between bodies in the interval from hundreds of nanometers to several {\AA}ngstr\"{o}ms    \cite{Reddy2015AIP,Kittel2017NatCommun,Reddy2017NatCommun}.
Generally, the results of these measurements turned out to be in good agreement with the predictions based on the fluctuation electrodynamics for a wide range of materials and geometries. However, there are still remain significant unresolved problems in understanding heat transfer between bodies in an extreme near field (gap size $<10$ nm)\cite{Kittel2017NatCommun,Reddy2017NatCommun}. In Refs.\cite{Kittel2017NatCommun,Reddy2017NatCommun} the heat flux  between a gold coated near-field scanning thermal microscope tip and a planar gold sample  in an extreme near field at nanometer distances of 0.2-7nm   was studied. It was found that  the experimental results can not be explained by the conventional theory of the radiative heat transfer  based on Rytov theory.  In particular, the heat transfer observed in Ref.\cite{Kittel2017NatCommun} for the separations from 1 to 10nm is orders of magnitude larger than the predictions of conventional Rytov theory and  its distance dependence is not well understood. These discrepancies stimulate  the search of the alternative channels of the heat transfer that can be activated in the extreme near-field. One of the obvious channels is related with ``phonon tunneling''  which stimulated active research of the phonon heat transfer in this region \cite{Volokitin2019JETPLett,Volokitin2020JPCM,Pendry2016PRB,Pendry2017Z.Nat,Persson2011JPCM,Volokitin2020JPCM_2D,Budaev2011APL,Joulain2014PRB,Sellan2012PRB,Esfarjani2015NCom,KapitzRes2016PRB,Roy2010PRL,Meltaus2010PRL,Henkel2019JOSA}. In Ref.\cite{Volokitin2019JETPLett,Volokitin2020JPCM} it was shown that the heat flux in the extreme near-field  can be strongly enhanced in presence of the potential difference between bodies. This enhancement is due to the electric field effect related with the fluctuating dipole moment induced on the surface by the potential difference. In this Rapid Communication  another mechanism,  related  with the fluctuating dipole moment of the electric double layer on metal surface,  is considered. In contrast with  the electric field effect, which is short range and operates for the separations of order $\sim 1$nm, the double layer effect is long range and operates for the separations $<10^2$nm.

\begin{figure}
\includegraphics[width=0.5\textwidth]{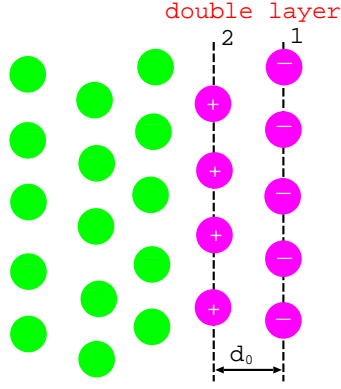}
\caption{ Scheme of an electric double layer on metal surface}
\label{Scheme}
\end{figure}  

The electric double layer on the metal surface is associated with a redistribution of the electron density, as a result of which the surface atomic layer 1 is charged negatively, and the near-surface layer 2 is positively charged (see Fig. \ref{Scheme}). The generic mechanism for such redistribution for metal surfaces is
the ``spill out'' of electron into vacuum. Deep in the metal the ion charges are compensating by negative electron density, but as the lattice abruptly terminates at the surface, electrons
tunnel out the solid over some small distance (Angstroms),
creating a negative sheet of charges in the surface atomic plane  and leaving a positive sheet of uncompensated metal ions in the
 sub-surface atomic planes\cite{Kahn2016Mater,Lang1971PRB}.The electric double layer can be approximately considered  as formed by two oppositely charged planes with a surface density $\pm\sigma_d$ (see Fig.\ref{Scheme}).This double layer  of charges creates a potential step $\Delta\varphi$ that increases the electron
potential just outside the surface, effectively raising work 
function\cite{Lang1971PRB}
\begin{equation}  
A=\Delta\varphi - \bar{\mu},
\label{A}
\end{equation}
where $\bar{\mu}=\mu-\bar{\varphi}$ is bulk chemical potential measured from the level of mean electrostatic potential.  

 The potential step depends on the condition of the surface, for example, the processing method and the degree of surface contamination. If the metal is a single crystal, then the surface potential relative to its inner part depends on the distance between the lattice planes parallel to the surface. A change in the crystallographic directions exposed on the surface of a pure polycrystalline metal leads to a change in the surface potential and the appearance of a surface charge density  $\sigma(\mathbf{x})$, for which the average value  $<\sigma(\mathbf{x})>=0$, but $<\sigma^2(\mathbf{x})>\neq 0$. This is named the ``patch effect'', which results in an electrostatic force of interaction between metal surfaces, which must be taken into account when measuring the Casimir force\cite{Speake2003PRL}. In addition to spatial fluctuations responsible for the ``patch effect'',  the dipole moment of the double layer will experience temporal fluctuations due to quantum and thermal fluctuations. Temporal fluctuations provide an additional contribution to the fluctuating electromagnetic field and associated fluctuation-induced electromagnetic phenomena, which include Casimir force and friction, and radiative heat transfer\cite{Volokitin2001PRB,Volokitin2007RMP,Volokitin2017Book}. In addition, the interaction of the electric field with the induced dipole moment of the double layer will lead to dissipation and to additional noncontact electrostatic friction between charged surfaces. A similar mechanism (elastic response of the first few layers to an external electric field) has been discussed in the 1980s to explain the 
anomalous infrared absorption of metallic nanoparticles\cite{Persson1984PRL,Apell1985SolidStateCommun}.

Consider two plates separated  by a vacuum gap  $d$.  In the near-field ($d<\lambda_T=c\hbar/k_BT$) the radiative heat transfer between them is dominated by the contribution from evanescent electromagnetic waves for which the  heat flux is determined by 
\cite{Volokitin2001PRB,Volokitin2007RMP,Volokitin2017Book}

\begin{equation}
J^{rad} =\frac{1}{\pi^2}\int_0^\infty d\omega\left[\Pi_1(\omega)-\Pi_2(\omega)\right]\int_0^{\infty} k_zdk_ze^{-2k_zd}
\left[\frac{
\mathrm{Im}R_{1p}(\omega,q)\mathrm{Im}R_{2p}(\omega,q) }{\mid 1-e^{-2
k_z d}R_{1p}(\omega,q)R_{2p}(\omega,q)\mid ^2}+(p\rightarrow s)\right],
\label{Heat}
\end{equation}
where
\[
\Pi_i(\omega)=\frac{\hbar \omega}{e^{\hbar\omega/k_BT_i}-1},
\]
 $R_p$ and $R_s$ are the reflection amplitudes for $p$ and $s$ polarized electromagnetic waves, $k_z=\sqrt{q^2-(\omega/c)^2}$, $q>\omega/c$ is the component of the wave vector parallel to the surface. In the presence of the electric double layer, the reflection amplitudes for the surface will not be determined by the Fresnel formulas, since the interaction of the electric field of the electromagnetic wave with the double layer will induce the surface dipole moment  $p_z=\alpha_{\perp}E_z^+$, where  $\alpha_{\perp}$ is the susceptibility of the double layer normal to the surface,  $E_z^+$ is the external normal component of the electric field on the surface.  In the presence of the surface dipole moment the reflection amplitude for the  $\textit{p}$ -polarized electromagnetic waves is determined by  \cite{Volokitin2019JETPLett}
\begin{equation}
R_p=\frac{i\varepsilon k_z  -k_z^{\prime} +4\pi iq^2\alpha_{\perp}\varepsilon}{i\varepsilon k_z  +k_z^{\prime} -4\pi iq^2\alpha_{\perp}\varepsilon}.
\label{rcp}
\end{equation}
The frequency domain equation of motion of plane 1 is  
\begin{equation}
(-\rho_d\omega^2-i\rho_d\omega\gamma_d+K)u_1-Ku_2=\sigma_dE_z,
\label{eqmotion}
\end{equation}
where  $\rho_d$ and $\gamma_d$  are  the density and damping constant for vibrations of  atomic plane 1, $K=Y/d_0$ is the elastic constant for the interplanar interaction, $Y$ is the Young modulus, $d_0$ is the interplane distance, $\sigma_d$ is the surface charge density of plane 1.   The displacement of surface 2 under action of stress $\sigma_2=K(u_1-u_2)-\sigma_dE_z$ 
is determined by 
\begin{equation}
u_2=M[K(u_1-u_2)-\sigma_dE_z],
\label{u2}
\end{equation}
where $M$ is the mechanical susceptibility that  determines  the surface displacement under 
the action of mechanical stress: $u=M\sigma^{ext}$. 
From Eqs.(\ref{eqmotion}) and (\ref{u2}) we get the dipole moment induced by the electric field in 
the double layer
\begin{equation}
p_z^{ind}= \sigma_d(u_1-u_2)=\alpha_dE_z,
\label{pz}
\end{equation}
where the susceptibility  of the double layer is given by
\begin{equation}
\alpha_{d} = \frac{\sigma_d^2[\omega_d^2-(\omega^2+i\omega\gamma_d)MK]}{K[\omega_d^2-(\omega^2+i\omega\gamma_d)(1+MK)]},
\label{alpha_d} 
\end{equation}
where $\omega_d=\sqrt{Y/\rho_dd_0}$. For gold   $Y=79$GPa, the separation between (111) planes $d_0=2.35${\AA}, the volume density $\rho=1.9280\cdot 10^4$kgm$^{-3}$, $\rho_d=\rho d_0=4.53\times10^{-6}$kgm$^{-2}$, $\omega_d=8.85\cdot 10^{12}$s$^{-1}$, the sound velocities for the longitudinal  and transverse  acoustic waves $c_l=3240$ms$^{-1}$and $c_t=1200$ms$^{-1}$ , the lattice constant $a=4.05\cdot 10^{-10}$m, the upper frequencies for the longitudinal and transverse and transverse acoustic waves $\omega_l^{max}=c_l\pi/a= 2.5\cdot 10^{13}$s$^{-1}$ and $\omega_t^{max}=c_t\pi/a= 9.3\cdot 10^{12}$s$^{-1}$. Because the resonant frequency $\omega_d$ lies inside the phonon band  of acoustic waves, for the mechanical susceptibility can be used expression obtained in a elastic continuum model \cite{Persson2001JPCM}
\begin{equation}
M=\frac{i}{\rho c_t^2}\left(\frac{\omega}{c_t}\right)^2\frac{p_l(q,\omega)}{S(q,\omega)},
\end{equation}
where
\[
S(q,\omega)=\left[\left(\frac{\omega}{c_t}\right)^2-2q^2\right]^2+4q^2p_tp_l,
\]
\[
p_t=\left[\left(\frac{\omega}{c_t}\right)^2-q^2+i0\right]^{1/2}, \,\,p_l=\left[\left(\frac{\omega}{c_l}\right)^2-q^2+i0\right]^{1/2}.
\]
The reflection amplitude for the  $\textit{s}$ -polarized electromagnetic waves is determined by
\begin{equation}
R_s=\frac{ik_z-k_z^{\prime}}{ik_z+k_z^{\prime}}.
\label{rcs}
\end{equation}
Another heat transfer mechanism is associated with electrostatic interaction between fluctuating surface dipole moments for which the heat flux\cite{Volokitin2019JETPLett,Volokitin2020JPCM}
\begin{equation}
J^{ph}=\frac{1}{\pi^2}\int_0^\infty d\omega\left[\Pi_1(\omega)-\Pi_2(\omega)\right]\int_0^\infty dq q
\frac{b^2\mathrm{Im}\alpha_1\mathrm{Im}\alpha_2 }{\mid (1-a\alpha_1)(1-a\alpha_2)-b^2\alpha_1\alpha_2
\mid^2},
\label{heatel}
\end{equation}
where 
\begin{equation}
a=4\pi q\mathrm{coth}qd,\,\, b=\frac{4\pi q}{\mathrm{sinh}qd}
\end{equation}
 For $q\gg |\varepsilon| \omega/c$ and $1/\varepsilon\ll 4\pi q\alpha \ll 1$ from Eq. (\ref{rcp}) $R_p\approx 1+8\pi q \alpha$ 
and from Eqs. (\ref{Heat}) and (\ref{heatel}) the contribution to the heat flux from the $p$-polarised waves
\begin{equation}
 J_p^{rad}\approx J^{ph}\approx J=16\int_0^\infty d\omega\left[\Pi_1(\omega)-\Pi_2(\omega)\right]\int_0^\infty dq q^3\frac{ \mathrm{Im}\alpha_1\mathrm{Im}\alpha_2}
{\mathrm{sinh}^2qd}
\label{heatapprox} 
\end{equation}

\begin{figure}
\includegraphics[width=0.5\textwidth]{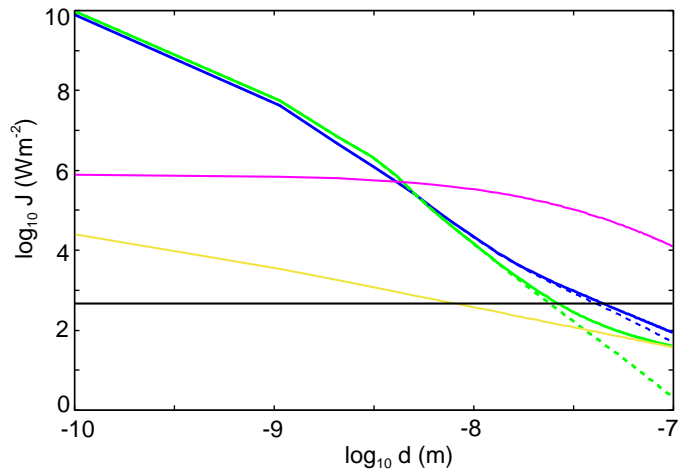}
\caption{The dependence of the heat flux between two plates of gold on the distance between them for different mechanisms. Temperature of one plate at $T = 300$K, and another at $T = 0$ K.  Solid  and dashed blue (green) lines  are  for the radiative heat flux  associated with $p$-polarized waves and  electrostatic phonon heat transfer, respectively, with damping constant  of the surface atomic layer vibrations $\gamma_d=0$ ($\gamma_d=10^{11}$s$^{-1}$).  Yellow  line shows   the radiative heat flux due to $p$ -polarized electromagnetic waves   without taking into account  the double layer effect. Pink line is  for the contribution from $s$ -polarized waves.  Black line for the radiative heat transfer associated with blackbody radiation. \label{Fig2}}
\end{figure}

Fig.\ref{Fig2} shows the dependence of the heat flux between two gold plates on the distance between them for different mechanisms at the temperature of one plate $T=300$K and  other at $T=0$K.  For gold dielectric function\cite{Chapuis2008PRB}
\begin{equation}
\varepsilon=1-\frac{\omega_p^2}{\omega^2+i\omega\nu},
\end{equation}
where  $\omega_p=1.71\times10^{16}$s$^{-1}$, $\nu=4.05\times 10^{13}$s$^{-1}$. In an extreme near field   according to the classical  Fresnel formulas the contribution from $p$-polarised waves is determined by \cite{Polder1971PRB}
\begin{equation}
S_p\approx \frac{0.2\hbar \nu^{0.5}}{cd\omega_p}\left(\frac{k_BT}{\hbar}\right)^{3.5}
\label{SpFr}
\end{equation}
and the $s$-wave contribution is distance independent
\begin{equation}
S_s\approx \frac{0.02\hbar\omega_p^2}{\nu c^2}\left(\frac{k_BT}{\hbar}\right)^3.
\label{SsFr}
\end{equation}
Such remarkable results are consequences  of singularities in the Fresnel transmission coefficients  at small $q$-wave vector for high conductivity materials like gold.     The  bulk chemical potential and work function  for (111) surface of gold     are   -1 and 5.31eV, respectively\cite{Lang1971PRB,Handbook2001}.  Thus from Eq.(\ref{A}) the potential step due to the double layer $\Delta\varphi\approx 4.3$eV.
Approximating the double layer by two opposite charged planes, the plane charge density can be estimated from the relation $4\pi\sigma_ded_0=\Delta\varphi$ where $d_0$ is the separation between planes. For gold the separation between (111) planes $d_0=2.35${\AA} thus $\sigma_d\approx 0.16$Cm$^{-2}$.  In contrast to the heat transfer between graphene and SiO$_2$ substrate  which is determined by damping of flexural vibrations of graphene sheet\cite{KapitzRes2016PRB}, the heat transfer between gold surfaces are not sensitive to the damping constant $\gamma_d$ (on Fig.\ref{Fig2} results for $\gamma_d=0$ and $\gamma_d=10^{11}$s$^{-1}$ are practically identical) and is mainly determined by excitation  of acoustic waves associated with coupling with double layer vibration with coupling constant $M$.  For $d>d_c=4.3$nm the heat flux is dominated by practically distance independent $s$-wave contribution. However for $d<d_c$ the heat flux is strongly enhanced  due to  $p$-wave contribution associated with contribution from the electrical double layer.  for which $S_p(d)\approx 10^{-10}/d^2$. The critical distance in heat transfer is determined by equation 
\begin{equation}
S_p(d_c)= S_s(d_c)
\label{dc}
\end{equation}
from where $d_c=4.3$nm.
 The heat flux in the tip-plane configuration can be obtained from the plane-plane configuration using proximity  (or Derjyaguin) approximation \cite{Parsegian2006}
\begin{equation}
P_{tip}=2 pi\int_0^Rd\rho\rho S(z(\rho))
\label{tipplane}
\end{equation}
where $R$ is the radius of the tip, $S(z)$ is the heat flux between two plates separated by distance $z(\rho)=d+R-\sqrt{R^2-\rho^2}$ denotes the tip-surface distance as a function of the distance $\rho$ from the tip symmetry. For $R=30$nm, $d=3.5\cdot 10^{-10}$, $n=2$ from Eq.(ref{tipplane}
\begin{equation}
P_{tip}^{max}=0.56\mu W
\label{Pmax}
\end{equation}
what agrees very well with the experimental data\cite{Kittel2017NatCommun}. Henkel \textit{et.al}\cite{Henkel2019JOSA} fit experimental results with equation
\begin{equation}
P\approx 2\pi R(d_c-d)\dot{q}.
\label{fit}
\end{equation}
However, in contrast to present theory, this fit does not provide microscopic explanation of physical origin of anomalously strong enhancement of heat transfer in an extreme near field.

\begin{figure}
\includegraphics[width=0.5\textwidth]{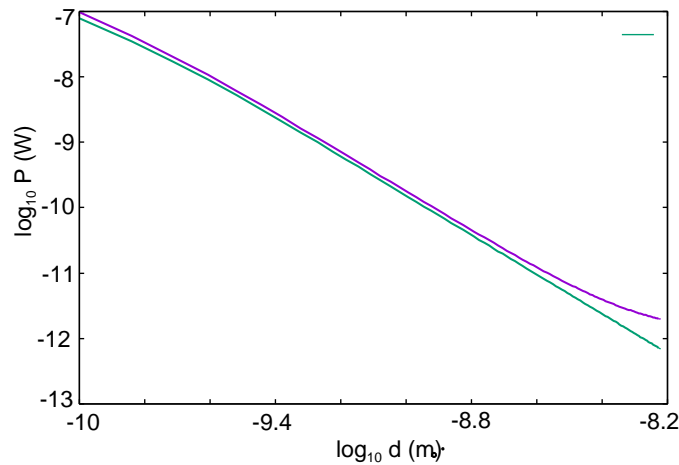}
\caption{The dependence of the heat flux between two gold samples on the distance between them in the tip-plate configuration. Temperatures of the tip and plate are   300 K and 0 K, respectively.  Blue and green lines  are  for the radiative heat flux  associated with $p$-polarized waves and  electrostatic phonon heat transfer, respectively.  Radius of the tip $R=30$nm.}  \label{Fig3}
\end{figure}

Fig. \ref{Fig3} shows dependence of the heat flux between gold samples in the tip-plate configuration. According to the conversional theory from Eq. (\ref{SpFr})

\begin{equation}
P_p\approx \frac{0.2\pi R\hbar \nu^{0.5}}{c\omega_p}\left(\frac{k_BT}{\hbar}\right)^{3.5}\mathrm{ln}(R/2d).
\label{PtFr}
\end{equation}
For $R\gg d$ this contribution weakly  depends on $d$ and at  $d=1${\AA} by five orders of magnitudes smaller than the contribution from electrical double layer. The maximum value $P^{max}\sim 1\mu W$ agrees with $P^{max}$ observed in Ref. \cite{Kittel2017NatCommun}. However distance dependence is more strong thus further work is needed to clarified details of this experiment. Similar situation occurrent for explanation of other phenomena related with fluctuation induced electromagnetic phenomena. Strong dependence of fluctuation induced phenomena related  with gold on structure of surface layer of gold was already noted in Ref.\cite{Volokitin2005PRL}. Experimentally it was demonstrated \cite{Stipe2001PRL} that between charged gold tip and and gold plate there is long range noncontact friction. However classical theory based on Fresnel formulas can not explained this results. Theory predicted friction which is by eleven orders magnitudes smaller experimental observation. In Ref.\cite{Volokitin2006PRB} it was proposed model which can explain puzzling experimental results. In this model it was proposed that surface is covered by incommensurate layer of adsorbates and within this model it was possible to explain distance dependence and magnitude for friction between gold surfaces. Thus to explain distance dependence and magnitude of heat flux in extreme near field more sophisticated model should be developed.

\textit{Conclusion.} 
The calculations of heat transfer between two plates of gold in an extreme near field were performed taking into account the electrical double layer effect. It was found that for $d<3$nm the double layer contribution exceeds the predictions of the conversional theory of the heat transfer by several orders of magnitudes.   The presented theory  predicts  anomalously large nano-scale heat transfer between metals as it was observed in Refs.\cite{Kittel2017NatCommun,Reddy2017NatCommun}.  Theory predicts the same strong enhancement of heat flux in an extreme  near field   as in experiment. The experimental results in Refs.\cite{Kittel2017NatCommun,Reddy2017NatCommun} were obtained  in the presence of the electrostatic potential difference between the STM tip and the sample  thus the electric field effect can also contribute to the heat flux. From the results presented above and in Refs.\cite{Volokitin2019JETPLett,Volokitin2020JPCM} follow that at the potentials difference of $\sim$1V   the electric field effect is negligible in comparison with the double-layer effect thus the electric field effect can be excluded from explaining the experimental results in Refs. \cite{Kittel2017NatCommun,Reddy2017NatCommun} for small bias voltage. However as it  was shown  in Messina et al. in arXiv:1810.02628 in
presence of a strong bias voltage the tunneling of electron dominates
in extreme near-field regime. The topic 
of heat transfer in presence of large bias voltage lies outside the scope of the present study. 
It follows from the results obtained that by intentionally changing the parameters of surface dipoles with the help of engineering, one can perform the  heat management  at the nanoscale, which is extremely important for nanotechnology. For example, adsorbates can be used to change the surface dipole moment and enhance or attenuate heat transfer\cite{Volokitin2004PRB}.   

\vskip 0.5cm

The reported study was funded by RFBR according to the research project N\textsuperscript{\underline{o}} 19-02-00453

\vskip 0.5cm

$^*$alevolokitin@yandex.ru

\end{document}